# SIMPLIFYING CREDIT SCORING RULES WITH LVQ+PSO

**Laura Cristina Lanzarini** and **Augusto Villa Monte**

Universidad Nacional de la Plata, Facultad de Informática, III-LIDI, La Plata, Argentina

**Aurelio F. Bariviera**

Department of Business, Universitat Rovira i Virgili, Reus, Spain,

**Patricia Jimbo Santana**

Facultad de Ciencias Administrativas, Universidad Central del Ecuador, Quito, Pichincha, Ecuador

**Abstract**

One of the key elements in the banking industry rely on the appropriate selection of customers. In order to manage credit risk, banks dedicate special efforts in order to classify customers according to their risk. The usual decision making process consists in gathering personal and financial information about the borrower. Processing this information can be time consuming, and presents some difficulties due to the heterogeneous structure of data. We offer in this paper an alternative method that is able to classify customers' profiles from numerical and nominal attributes. The key feature of our method, called LVQ+PSO, is the finding of a reduced set of classifying rules. This is possible, due to the combination of a competitive neural network with an optimization technique. These rules constitute a predictive model for credit risk approval. The reduced quantity of rules makes this method not only useful for credit officers aiming to make quick decisions about granting a credit, but also could act as borrower's self selection. Our method was applied to an actual database of a credit consumer financial institution in Ecuador. We obtain very satisfactory results. Future research lines are exposed.

1. Introduction

The economic development in the last sixty years was accompanied by an extension and popularization of the financial services. In fact, consumer lending gives the opportunity to a large part of the population of many countries to obtain some goods and services now, deferring the payment sometime in the future. This sort of "democratization" in consumption poses a challenge to financial institution. Whereas mortgage lending applications, due to its comparatively reduced number of borrowers, can be decided at a slower pace, consumer lending needs faster decision procedures. Borrowers want small credits for buying home equipment, a car, a trip, etc. They are eager of a quick answer. Financial institutions want to find the appropriate rules in order to approve credit application only to good borrowers, i.e. those who pay back their financial commitments. From the point of view of the borrowers, they want to receive a positive answer to their applications.

Financial institutions typically ask exhaustive information about the potential client: age, marital status, salary, other debts, job type, etc. This information is gathered in order to be analyzed, using some decision model. The result of this analysis is either to grant or reject the credit.

The increasing number of applicants and data raises the necessity for suitable techniques that deals with the complexity of this multidimensional problem. Precisely, the area known as data mining can shed light on this kind of situations. Lessmann et al (2015) affirm that the business value of accurate prediction relies on its relation with the firm profit equation.

Data mining comprises a set of techniques that are able to model available information. One of the most important stages in the process is knowledge discovery. It is characterized by obtaining new and useful information without assuming prior hypothesis. One of the preferred techniques by decision makers is the association rule.

An example of association rule is an expression: IF condition1 THEN condition2, where both conditions are conjunctions of propositions of the form (attribute = value) and whose solely restriction is that attributes in the antecedent must no be present in the consequent. When a set of association rules presents in the consequent the same attribute it is called a set of classification rules (Witten et al. 2011, Hernández and Ramírez, 2004).

The aim of this paper is twofold. On one hand we benchmark a method for obtaining classification rules that combines a neural network with an optimization technique, against two well-known classification methods. On the other hand, we show that the solution provided is very intuitive and simple, due to the reduced number of rules required for the decision.

A reduced set of rules improves the transparency in the decision making process of the financial institutions.

The rest of the paper is structured as follows. Section 2 briefly discusses relevant literature on credit risk. Section 3 describes the neural network, metaheuristics, and the proposed method. Section 4 describes data and presents results of a true empirical application and section 5 draws the main implications of our proposal.

## 2. Related work

The interest in studying business risk can be traced back to FitzPatrick (1932), who wrote one of the earliest papers in bankruptcy prediction, using 13 accounting ratios calculated for 40 firms during three years. In the 1960s, the development of the capital markets in the United States, showed the necessity for more scientific models to assess economic corporate strength. Consequently, the first z-score model by Altman (1968) was developed. At that time, the main concern of banks was to classify corporations according to their credit risk, since they were the main clients. However, in the last decades, there has been an increase in consumer credit. Retail banking is a growing industry. Not only there has been a boom in credit card memberships, specially in emerging economies, but also an increase in small consumption credits. For example, it is very common in emerging economies that families buy home appliances instalments. In those countries, it is usual the association of a home appliance shop with a financial institution, in order to provide customers with quick decision credit line facilities. The existence of such financial instrument aids to boost sales. This association generates a conflict of interests. On one hand, the home appliance shop wants to sell products to all customers. Therefore, it is in its best interest to promote a generous credit policy. On the other hand, the financial institution wants to maximize the revenue from credits, which lead to a strict surveillance of loan losses. Having a fair and transparent credit granting policy favors good business relationship between home appliances shops and financial institutions. One way of developing such policy is to construct objective rules in order to decide to grant or deny a credit application.

There are several methods to construct rules in order to evaluate the creditworthiness of credit applicants. The earliest methods were developed based in a discriminant analysis similar to Altman (1968). However, computational intelligent techniques produce better results. These techniques, without being exhaustive, include artificial neural networks, fuzzy set theory, decision trees, support vector machines, genetic algorithms, among others. Artificial neural networks is a family of neural networks with different architectures. These architectures includes popular models such as back propagation networks, self-organizing maps and learning vector quantization. Fuzzy set theory, developed from the seminal paper by Zadeh (1965) results very useful in cases such as credit classification, where boundaries are not crisp defined. Decision trees transform data in a tree-shape structure of leaf and decision nodes, and the goal is to test attributes to each branch of the tree, that constitutes a class. Support vector machines search an optimal hyperplane in order to generate a binary classification, maximizing the margin of separation between classes. Genetic algorithms are a set of methods to optimized problems, based on the evolutionary idea of natural selection. Hand and Henley (1997) highlight the difficulty in discovering new statistical techniques in this field, due to the need for confidentiality. Better techniques provide a competitive advantage to financial institutions, and are not willing to disclose such discovery. Freitas (2003) discusses the use of genetic algorithms in data mining and classification problems. Wang et al. (2007) propose a classification rule mining algorithm based on particle swarm optimization. Lessmann et al (2015) find that Artificial Neural Networks perform better than Extreme Learning Machine. Abid et al. (2016) use logistic regression and discriminant analysis in order to separate "good" and "bad" borrowers from a database of a commercial Tunisian bank for the period 2010-2012. For a more detailed and recent review of both traditional statistical models and intelligent methods for financial distress forecasting, we refer to Chen et al. (2016) and references therein.

If the goal is to obtain association rules, the a priori method (Agrawal and Srikant, 1994) or some of its variants could be used. This method identifies the most common sets of attributes and then combines them to get the rules. There are variants of the a priori method, are usually oriented reduce computation time.

Under the topic classification rules, the literature contains various construction methods based on trees such as C4.5 (Quinlan, 1993) or clipped trees as the PART method (Frank and Witten, 1998). In both cases, the key is to get a set of rules that covers the examples fulfilling a preset error bound. The methods of construction rules from trees are partitives and are based on different attributes' metrics to assess its ability to cover the error bound.

3. Methodology

This paper presents a hybrid approach based on Particle Swarm Optimization (PSO) to determine the rules. There are methods of obtaining rules using PSO (Wang et al., 2007). However, when operating with nominal attributes, a sufficient number of examples to cover all areas of the search space is required. If this situation is not feasible, its consequence is a poor initialization of the population, leading to premature convergence. As a way to bypass this problem, while reducing the turnaround time, is to obtain the initial state from a competitive LVQ neural network (Learning Vector Quantization). There is some literature that uses PSO as a means to determine the optimal quantity of competitive neurons to be used in the network, such as Hung and Huang (2010). This is not the purpose of this paper since the LVQ network we used, although it is previously dimensioned, it could estimate the number of neurons to be used for each class based on the proportion of examples in the training set.

### 3.1. Learning Vector Quantization (LVQ)

Learning Vector Quantization (LVQ) is a supervised classification algorithm based on centroids or prototypes (Kohonen, 1990). It can be interpreted as a three-layer competitive neural network. The first layer is only an input layer. The second layer is where the competition takes place. The third layer performs the classification. Each neuron in the competitive layer has an associated numerical vector of the same dimension as the input examples and a label indicating the class they will represent. These vectors are the ones that, at the end of the adaptive process, will contain information about the classification prototypes or centroids. There are different versions of the training algorithm. We will describe the one used in this article.

When starting the algorithm, some amount K centroids should be indicated. This allows defining the network architecture, given that number of inputs and outputs are defined by the problem.

Centroids are initialized taking K random examples. Then, examples are entered one at a time in order to adapt the position of the centroids. In order to do this, the closest centroid to the example is determined, using a preset distance measure. Since this is a supervised process, it is possible to determine whether the example and the centroid correspond to the same class. If the centroid and the example belong to the same class, the centroid to moved closer to the example with the aim of strengthening the representation. Conversely, if the classes are different, is the centroid is moved away from the example. These movements are performed using a factor or adaptation rate. This process is repeated either until changes are less than a pre-set threshold or until the examples are identified with the same centroids in two consecutive iterations, whichever comes first.

In the implementation used in this article, we also examine the second nearest centroid and, in case it belongs to a different class of the example and be at a distance of less than 1.2 times the distance to the first centroid, it is moved away. Several variants of LVQ can be consulted in Kohonen et al. (2001).

### 3.2. Obtaining classification rules with particle swarm optimization (PSO)

Particle Swarm Optimization (PSO) is a population-based metaheuristic proposed by Kennedy and Eberhart (1995). In it, each individual in the population (particle) represents a possible solution to the problem and adapts following three factors: knowledge on the environment (fitness value), historical knowledge or past experience (memory) and historical knowledge or previous experiences of individuals located in its neighborhood (social knowledge).

PSO was originally defined to work on continuous spaces. In order to operate with it on a discrete space it is necessary to take into account some precautions. Kennedy and Eberhart (1997) defined a binary version of PSO method. One of the central problems of the latter method is its difficulty changing from 0 to 1 and from 1 to 0 once it has stabilized. This has led to different versions of binary PSO, looking to improve the exploratory capacity. In particular, this work will use a variant defined by Lanzarini et al. (2011).

Obtaining classification rules using PSO, when operating on nominal and numeric attributes, requires a combination of the methods mentioned above. This is so, because it is necessary to say which attributes will be part of antecedent and what value or range of values it may take (a combination of discrete and continuous spaces).

Since it is a population technique, it should be analyzed the required information in each individual of the population. A decision between representing a single rule or the full rules set per individual should be made. At the same time, the representation scheme of each rule should be chosen. Tanking into account the aim of this work, we follow the Iterative Rule Learning (IRL) approach developed by Venturini (1993), in which each individual represents a single rule and the solution is constructed from the best individuals obtained in a sequence of executions. Consequently, using this approach implies that the population technique be applied iteratively until the desired coverage, obtaining a single rule for each iteration: the best individual of the population. It has also been decided to use a fixed length representation where only the antecedent of the rule is coded and given this approach, an iterative process will associate all individuals in the population with a default class, which does not require coding the consequent.

Regarding the fitness of each individual, it depends on two things: firstly the importance of the rule that represents (based on its support and confidence) and secondly the size (proportion of attributes used in the antecedent relative to the total number of attributes).

A detailed description on the application of PSO for obtaining classification rules is in Lanzarini *et al.* (2015b).

### 3.3. LVQ+PSO. Proposed method for obtaining rules

Rules are obtained through an iterative process that analyzes examples not covered in each class, beginning by the more populated classes. Whenever a rule is obtained, examples covered by such rule are removed from the set of input data. The process continues until covering all examples, or until the amount of uncovered examples in each class examples is either below the respective minimum established support or until they the maximum number of attempts to obtain a rule have been reached. It is important to note that, since examples are removed from the set of input data once they are covered by the rules, they constitute a classification. This is to say that, in order to classify a new example, rules must be applied in the order in which they were obtained and the example will be classified according to the corresponding class of the consequent of the first rule whose antecedent verifies for the example under examination.

Before starting the iterative process of obtaining rules, the method starts with the supervised training of a LVQ neural network, using the full set of examples and the algorithm described in Section 2. The goal of this step is to identify the most promising areas of space search.

Since neural networks operate only with numeric data, nominal attributes are represented by a dummy coding using as many binary digits as the different options of the nominal attribute. In addition, before starting the training, each numeric attribute is linearly scaled in the interval [0, 1]. The similarity measure used is the Euclidean distance. Once training is complete, each centroid will contain approximately the average of the examples it represents.

In order to obtain each of the rules, it is determined firstly, which is corresponding class of the consequent. With the aim of obtaining rules with high support, the proposed method analyzes the classes having a greater number of uncovered examples. The minimum support that a rule must meet is proportional to the amount of non covered examples of the class by the time that was obtained. In other words, the minimum support required for each class decreases along iterations, as examples of the corresponding class are covered. Thus, it is expected that the first rules have more support than the last rules.

Once the class is selected, the consequent is determined by the rule. In order to obtain the antecedent, a swarm population will be optimized, using the algorithm described in Section 3, initialized with the information of all centroids able to represent a minimum number of examples from the selected class and its immediate neighbors. The information of the centroid is used to determine vector veloc2, described in Section 3. If this is a nominal attribute, centroid information is linearly scaled to the interval [*lowerbound2$_j$, upperbound2$_j$*]. However, if it is a numeric attribute the value to be scaled is (*1-1.5 \* deviation$_j$*) being *deviation$_j$* the j-*th* dimension of the deviation of the examples represented by the centroid. In both cases, it is intended to operate with a value between 0 and 1 that measures the degree of participation of the attribute (if numeric) or attribute value (if nominal) in building the antecedent of the rule. In the case of nominal attributes, it is clear that the average indicates the ratio of elements represented by the centroid that match the same value. However, when it is numeric, this ratio is not present in the centroid but the deviation of the examples (considering a specific dimension). If the deviation in a certain dimension is zero, all examples coincide in the value of the centroid, but if it is too large, it should be understood that it is not representative of the group. Therefore, it would not be appropriate to include it in the antecedent of the rule. If deviation is large, using ($1- 1.5 * deviation_j$), the speed value *veloc2* (argument of the sigmoid function) will be lower and the probability that the attribute be used is reduced. In all cases the speed veloc1 is initialized randomly in [$lowerbound1_j$, $upperbound1_j$]. Figure 1 shows the pseudocode of the proposed method.

```
Train LVQ network using all training examples
Compute the minimum support for each class
While (the end criteria is not reached)
      Choose the class with largest number of non
      covered examples
      Construct a reduced population of the
      individuals, based on centroids
      Evolve the population using PSE according
      section 4
      Obtain the best population rule
      If (the rule fulfils with support and
      confidence required) then
         Add the rule to the set of rules
      Consider as covered the examples correctly
      classified by the previous rule
      Recalculate the minimum support for this
      class
    End if
End while
```

**Fig. 1 Pseudocode of the proposed method.**

4. **Data and Results**

We test our method in real consumer credit records of a savings and credit institution of Ecuador, which generously provided data. The data comprises credit operations between January 2011 until August 2015, with the following attributes: status; date of application; branch; province; requested amount; authorized amount; purpose of the credit; cash, bank accounts, investments, other assets, liabilities and salary of the applicant; date of verification of information; date of authorization; approval/denial date; cash, bank accounts, investments, other assets, liabilities and salary of the applicants' partner. In case, the applicant is a small business data requested are revenues and expenses of the business. The 'status' variable correspond to the situation of the credit. Applications can be denied or accepted. In case of being accepted, the status is classified between credits that were duly repaid and those with some delay in the payback. In turn, overdue loans are classified, according to the credit procedures manual between those with less than 90 days overdue, and those with more than 90 days overdue (initiation of legal actions).

Using the data described above, we compare the performance of the proposed method, LVQ + PSO, *vis-à-vis* C4.5 methods defined by Quinlan (1993) and PART defined by Witten et al. (2011). Both alternative methods allow classification rules. C4.5 is a pruned tree whose branches are mutually exclusive and allow classifying examples. PART gives as a result a list of rules equivalent to those generated by the proposed classification method, but in a deterministic way. PART operation is based on the construction of partial trees. Each tree is created in a similar manner to that proposed for C4.5 but during the process construction errors of each branch are calculated. These errors allow the selection of the most suitable combinations of attributes. For a detailed description of the method see PART [1].

We performed 10 independent runs of each method. For LVQ+PSO, we use a LVQ network of 30 neurons distributed between classes in proportion to the examples used.

PART method was executed with a confidence factor of 0.3 for the pruned tree. For other parameters default values were used.

Tables 1 summarizes the results obtained by applying the three methods. In each case was considered not only the accuracy of coverage of the rule set, but also the "transparency" of the obtained model. This "transparency" is reflected in the average number of rules obtained and the average number of terms used to form the antecedent. We would like to highlight that, as we said in the introduction, the proposed method is simple. This simplicity gives the general manager of a financial institution a clear profile of the "good customer". This situation could benefit the firm not only through a reduction of the default risk, but also to help to find the right customers in the future, through marketing campaigns.

**Table 1**. Prediction Results of the proposed and benchmark methods

| Method | Prediction | Deny | Accept | Type I error | Precision | # rules | | Antecedent | |
|---|---|---|---|---|---|---|---|---|---|
| C4.5 | Deny | 1422.60 | 244.18 | 0.11 | 81.05 | 114.16 | 8.66 | 9.70 | 0.19 |
|  | Accept | 181.61 | 398.61 |  |  |  |  |  |  |
| PART | Deny | 1407.15 | 238.58 | 0.11 | 80.61 | 41.97 | 1.85 | 4.71 | 0.11 |
|  | Accept | 197.04 | 404.23 |  |  |  |  |  |  |
| LvqPSOVar | Deny | 1450.26 | 314.73 | 0.14 | 79.20 | 3.12 | 0.09 | 2.54 | 0.17 |
|  | Accept | 152.75 | 329.26 |  |  |  |  |  |  |

In a previous work, Lanzarini et al. (2015a) showed using public databases that LVQ+PSO achieves higher accuracy higher than PART but equivalent to that achieved by C4.5 method.

In our case, even though the precision of our method is slightly lower than the benchmark models, the number of rules is significantly lower. In fact, our method needs less than 3% of the rules of C4.5 and 7.5% of the rules of PART. The antecedent is also shorter in our method than in the benchmark models. Consequently, we believe that our model is suitable for credit scoring. In fact, it is much more simple and straightforward to understand by the decision maker. Considering a trade-off between number of rules and precision/Type I error, we believe that our model is quite acceptable, taking into account that it provides understandable information to managers, in order to target the right potential customers in the future.

## 5. Conclusions

We introduce a competing method for credit scoring using a variation of binary PSO, whose population is initialized with information from the centroids of a network previously trained LVQ neural network. The advantage of this dual treatment is that it allows to deal with numerical and nominal attributes, as it is the usual case in credit applications.

We test our model on actual credit operations from an important retail credit institution from Ecuador. Results show clearly that the LVQ + PSO method obtains a simpler model. It uses about 7.5% of the quantity of rules generated by PART and 3% of the rules needed by C4.5, with an antecedent formed by few conditions and slightly worse accuracy.

In spite of the fact that conducted tests showed no evidence of dependence between results and the initial size of the LVQ network, it is considered desirable to repeat the measurements using an LVQ network of minimum size and a version of variable population PSO to adequately explore the solution space in the future.

Finally, we would like to highlight that goal of our method is to achieve an intuitive model for credit scoring with a comparable accuracy to popular benchmark models. Our results suggest that the simplification of decision rules generates transparency in credit scoring, which could improve the reputation of financial institutions.

**References**

Abid, L., Masmoudi, A. & Zouari-Ghorbel, S., 2016. The Consumer Loan's Payment Default Predictive Model: an Application of the Logistic Regression and the Discriminant Analysis in a Tunisian Commercial Bank. *Journal of the Knowledge Economy*, pp.1–15. Available at: http://dx.doi.org/10.1007/s13132-016-0382-8.

Agrawal, R.,Srikant, R., 1994. Fast algorithms for mining association rules in large databases. In: Proceedings of the 20th International Conference on Very Large Data Bases, VLDB '94, pp. 487–499. Morgan Kaufmann Publishers Inc., San Francisco.

Altman, E.I., 1968. Financial ratios, discriminant analysis and the prediction of corporate bankruptcy. *The Journal of Finance*, 23(4), pp.589–609. Available at: http://doi.wiley.com/10.1111/j.1540-6261.1968.tb00843.x.

Chen, N., Ribeiro, B. & Chen, A., 2016. Financial credit risk assessment: a recent review. *Artificial Intelligence Review*, 45(1), pp.1–23. Available at: http://dx.doi.org/10.1007/s10462-015-9434-x.

FitzPatrick, P.J., (1932). A comparison of the ratios of successful industrial enterprises with those of failed companies. *The Certified Public Accountant*, Oct., Nov., Dec.

Frank, E., Witten, I. H., 1998. Generating accurate rule sets without global optimization. In: Proceedings of the Fifteenth International Conference on Machine Learning, ICML '98., pp. 144–151. Morgan Kaufmann Publishers Inc., San Francisco.

Freitas, A.A., 2003. A Survey of Evolutionary Algorithms for Data Mining and Knowledge Discovery. In A. Ghosh & S. Tsutsui, eds. *Advances in Evolutionary Computing: Theory and Applications*. Berlin, Heidelberg: Springer Berlin Heidelberg, pp. 819–845. Available at: http://dx.doi.org/10.1007/978-3-642-18965-4_33.

Hand, D.J., Henley, W.E., 1997. Statistical Classification Methods in Consumer Credit Scoring: A Review. *Journal of the Royal Statistical Society. Series A (Statistics in Society)*, 160(3), pp.523–541. Available at: http://www.jstor.org/stable/2983268.

Hernández Orallo, J., Ramírez Quintana, M.J., Ferri Ramírez, C., 2004. *Introducción a la Minería de Datos*. 1ra Edición. Pearson.

Hung, C. & Huang, L., 2010. Extracting Rules from Optimal Clusters of Self-Organizing Maps. In *Second International Conference on Computer Modeling and Simulation. ICCMS '10.* pp. 382–386.

Kennedy, J. & Eberhart, R., 1995. Particle swarm optimization. In*, Proceedings of IEEE International Conference on Neural Networks*. pp. 1942–1948 vol.4.

Kennedy, J. & Eberhart, R.C., 1997. A discrete binary version of the particle swarm algorithm. In *IEEE International Conference on Systems, Man, and Cybernetics, 5,* pp. 4104–4108.

Kohonen, T., 1990. The self-organizing map. *Proceedings of the IEEE*, 78(9), pp.1464–1480.

Kohonen, T., Schroeder, M.R., Huang, T.S. (Eds.), 2001. Self-Organizing Maps, 3rd ed. Springer-Verlag, New York.

Lanzarini, L., López, J., Maulini, J.A., De Giusti, A., 2011. A New Binary PSO with Velocity Control. In Y. Tan et al., eds. *Advances in Swarm Intelligence: Second International Conference, ICSI 2011, Chongqing, China, June 12-15, 2011, Proceedings, Part I*. Berlin, Heidelberg: Springer Berlin Heidelberg, pp. 111–119. Available at: http://dx.doi.org/10.1007/978-3-642-21515-5_14.

Lanzarini, L. Villa Monte, A., Bariviera, A.F., Jimbo Santana, P., 2015a. Obtaining Classification Rules Using LVQ+PSO: An Application to Credit Risk. In J. Gil-Aluja et al., eds. *Scientific Methods for the Treatment of Uncertainty in Social Sciences*. Advances in Intelligent Systems and Computing. Cham: Springer International Publishing, pp. 383–391. Available at: http://link.springer.com/10.1007/978-3-319-19704-3.

Lanzarini, L., Villa-Monte, A., Ronchetti, F., 2015b. SOM+PSO. A Novel Method to Obtain Classification Rules. Journal of Computer Science & Technology (JCS&T), 15(1), pp. 15-22.

Lessmann, S., Baesens, B., Seow, H-S., Thomas, L.C., 2015. Benchmarking state-of-the-art classification algorithms for credit scoring: An update of research. *European Journal of Operational Research*, 247(1), pp.124–136. Available at: http://www.sciencedirect.com/science/article/pii/S0377221715004208.

Quinlan, J.R., 1993. *C4.5: programs for machine learning*, Morgan Kaufmann Publishers.

Venturini, G., 1993. SIA: A supervised inductive algorithm with genetic search for learning attributes based concepts. In P. B. Brazdil, ed. *Machine Learning: ECML-93: European Conference on Machine Learning Vienna, Austria, April 5--7, 1993 Proceedings*. Berlin, Heidelberg: Springer Berlin Heidelberg, pp. 280–296. Available at: http://dx.doi.org/10.1007/3-540-56602-3_142.

Wang, Z., Sun, X. & Zhang, D., 2007. A PSO-Based Classification Rule Mining Algorithm. In D.-S. Huang, L. Heutte, & M. Loog, eds. *Advanced Intelligent Computing Theories and Applications. With Aspects of Artificial Intelligence: Third International Conference on Intelligent Computing, ICIC 2007, Qingdao, China, August 21-24, 2007. Proceedings*. Berlin, Heidelberg: Springer Berlin Heidelberg, pp. 377–384. Available at: http://dx.doi.org/10.1007/978-3-540-74205-0_42.

Witten, I.H., Frank, E. & Hall, M.A., 2011. *Data Mining Practical Machine Learning Tools and Techniques* 3rd. ed., San Francisco, CA: Morgan Kaufmann Publishers Inc.

Zadeh, L.A., 1965. Fuzzy sets. *Information and Control*, 8(3), pp.338–353. Available at: http://linkinghub.elsevier.com/retrieve/pii/S001999586590241X.